\documentclass[english,prl,twocolumn]{revtex4}
\usepackage{lmodern}
\usepackage{lmodern}
\usepackage[T1]{fontenc}
\usepackage[latin9]{inputenc}
\usepackage{textcomp}
\usepackage{amsmath}
\usepackage{graphicx}
\usepackage{esint}
\usepackage{upgreek}

\makeatletter

\newcommand{\lyxmathsym}[1]{\ifmmode\begingroup\def\b@ld{bold}
  \text{\ifx\math@version\b@ld\bfseries\fi#1}\endgroup\else#1\fi}

\@ifundefined{textcolor}{}
{%
 \definecolor{BLACK}{gray}{0}
 \definecolor{WHITE}{gray}{1}
 \definecolor{RED}{rgb}{1,0,0}
 \definecolor{GREEN}{rgb}{0,1,0}
 \definecolor{BLUE}{rgb}{0,0,1}
 \definecolor{CYAN}{cmyk}{1,0,0,0}
 \definecolor{MAGENTA}{cmyk}{0,1,0,0}
 \definecolor{YELLOW}{cmyk}{0,0,1,0}
}

\usepackage{verbatim}
\usepackage{babel}

\usepackage{babel}

\makeatother

\usepackage{babel}
\begin{document}

\title{Current crowding mediated large contact noise in graphene field-effect
transistors}

\author{Paritosh Karnatak$^{1}$\footnotemark[1]\footnotemark[2], T. Phanindra
Sai$^{1}$\footnote{Equal Contributions}\footnote{email: phanindra.s@gmail.com, paritosh@physics.iisc.ernet.in}, Srijit Goswami$^{1}$\footnotemark[1]\footnote{Current affiliation: QuTech and Kavli Institute of Nanoscience, Delft University of Technology, 2600 GA Delft, The Netherlands.}, Subhamoy Ghatak$^{1}$, Sanjeev Kaushal$^{3}$, Arindam Ghosh$^{1,2}$}

\affiliation{$^{1}$Department of Physics, Indian Institute of Science, Bangalore
560 012, India. }

\affiliation{$^{2}$ Centre for Nano Science and Engineering, Indian Institute
of Science, Bangalore 560 012, India.}

\affiliation{$^{3}$Tokyo Electron Ltd., Akasaka Biz Tower, 3-1 Akasaka 5-Chome,
Minato-ku, Tokyo 107-6325, Japan.}
\begin{abstract}
The impact of the intrinsic time-dependent fluctuations in the electrical
resistance at the graphene-metal interface or the contact noise,
on the performance of graphene field effect transistors, can
be as adverse as the contact resistance itself, but remains largely
unexplored. Here we have investigated the contact noise in graphene
field effect transistors of varying device geometry and contact configuration, with 
carrier mobility ranging from 5,000~cm$^{2}$V$^{-1}$s$^{-1}$ to 80,000~cm$^{2}$V$^{-1}$s$^{-1}$.
Our phenomenological model for contact noise due to current crowding
in purely two dimensional conductors, confirms that the contacts dominate
the measured resistance noise in all graphene field effect transistors in
the two-probe or invasive four probe configurations, and surprisingly,
also in nearly noninvasive four probe (Hall bar) configuration in
the high mobility devices. The microscopic origin of contact noise
is directly linked to the fluctuating electrostatic environment of
the metal-channel interface, which could be generic to two dimensional material-based
electronic devices. 
\end{abstract}


\maketitle

\section*{Introduction}
The wide spectrum of layered two-dimensional materials provides the
opportunity to create ultimately thin devices with functionalities
that cannot be achieved with standard semiconductors. The simplest
of such devices is the field effect transistor (FET). There are several
factors which determine the performance of an FET, key among them
being the dielectric environment, quality of the metal-semiconductor
contact, and the level of low-frequency $1/f$ noise. Over the past
few years there has been tremendous progress in creating high mobility,
atomically thin FETs through a combination of low resistance ohmic
contacts~\cite{Goddard_end_contact,SmithAcsNano2013,graphene_edge_contact,edge_contact2}
and strategies for encapsulation~\cite{graphene_edge_contact} of
the active channel. However, there exists no consensus on the factors
which determine the magnitude of the $1/f$ noise, which is known
to degrade the performance of amplifiers, or introduce phase noise/jitter
in high frequency oscillators and converters~\cite{phase_noise}.
Noise is especially detrimental to the performance of nanoscale
devices and may cause variability even in ballistic transistor channels,
where it has been suggested to arise from slow fluctuations in the
electrostatic environment of the metal-semiconductor contacts~\cite{TersoffNanoLett2007}.

Even for the widely studied graphene FET it is still unclear what
the dominant contribution is to the $1/f$ noise. Conflicting claims
exist, where some studies attribute the $1/f$-noise in graphene transistors
primarily to noise generated within the channel region~\cite{RumyantsevJoP2010,PalACSNano2011,HellerNanoLett2010},
whereas other investigations indicate a strong contribution from the
contacts~\cite{graded_graphene,LiuAPL2009,Suspended_contact_noise}.
This distinction has remained elusive to existing studies~\cite{LinNanoLett2008,PalPRL2009,HellerNanoLett2010,RumyantsevJoP2010,M-shape_noise,PalACSNano2011,graded_graphene,LiuAPL2009,Suspended_contact_noise,KaverzinPRB2012,BalandinNatNano2013,pellegrini_noise,edge_contact_noise,NoiseGBN,ChandanNL}
due to the lack of a microscopic understanding of how processes characteristic
to the metal-graphene junctions, in particular the current crowding
effect~\cite{current_crowd,vandammecurrentcrowd,Grosse_current_crowding,NagashioAPL2010,P_current_crowding,MoS2_current_crowding},
impact the nature and magnitude of $1/f$ noise.

Fundamentally, current crowding is an unavoidable consequence of resistivity
mismatch at the metal-semiconductor junction, where the injection
and/or scattering of charge carriers between the semiconductor and
the metal contact is restricted only close to the edge of the contact,
over the charge transfer length $L_{{\rm T}}$~\cite{current_crowd,berger1972contact}.
Photocurrent measurements~\cite{Photo_contact,photocurrentxia2009,photoscanning}
and Kelvin probe microscopy~\cite{Kelvin_probe,Kelvin_probe1} at
the graphene-metal interface have already indicated the presence of
current crowding with $L_{{\rm T}}\sim0.1-1~\upmu$m~\cite{SongCarbonLett2013,Grosse_current_crowding,NagashioAPL2010}.
Restricting the effective current injection area leads to greater
impact of local disorder kinetics, and hence larger $1/f$ noise~\cite{vandamme_contact_noise}.
For graphene-metal interfaces, the scenario is more complex than a typical metal-semiconductor junction, since
it is known that metals such as Cr, Pd, and Ti react to form metal
carbides with graphene, altering the structural properties and causing
strong modifications in its energy band dispersion~\cite{GongAcsNano2014,SongCarbonLett2013}.
While it is clear that current crowding and the characteristics of
the metal-graphene junction directly influence the contact resistance~\cite{GongAcsNano2014,NagashioAPL2010,Photo_contact,HuardPRB2008,XiaNatNano2011,WFpinning,SongCarbonLett2013,Rc_Mayank,Contact_resistance_2part,independent_contactR},
how these factors impact the noise originating at the contacts (contact
noise) is still not known. 

In this work we study a series of graphene FETs with different mobilities,
substrates and contacting configurations to demonstrate that electrical
noise at the metal-graphene junction can be the dominant source of
$1/f$ noise in graphene FETs, especially for invasive contacting
geometry, where the probe contacts lie directly in the path of the
current flow. The contact noise was found to scale as $R_{{\rm c}}^{4}$,
where $R_{{\rm c}}$ is the contact resistance, in all devices and at all
temperatures. While the noise magnitude is determined by the fluctuating
charge trap potential at the oxide substrate underneath the metal
contacts, a simple phenomenological model unambiguously attributes
the scaling to the current crowding effect at the metal-graphene junction.
In view of the recent observations of contact noise~\cite{ghatak_noise_MOS2,paul2015percolative}
and current crowding effect in molybdenum disulphide (MoS$_{2}$)
and black phosphorus FETs~\cite{MoS2_current_crowding,P_current_crowding},
many of the results and concepts developed in this paper can be extended
to other members of 2D semiconductor family as well. 

\begin{figure*}[t]
\includegraphics[width=2\columnwidth]{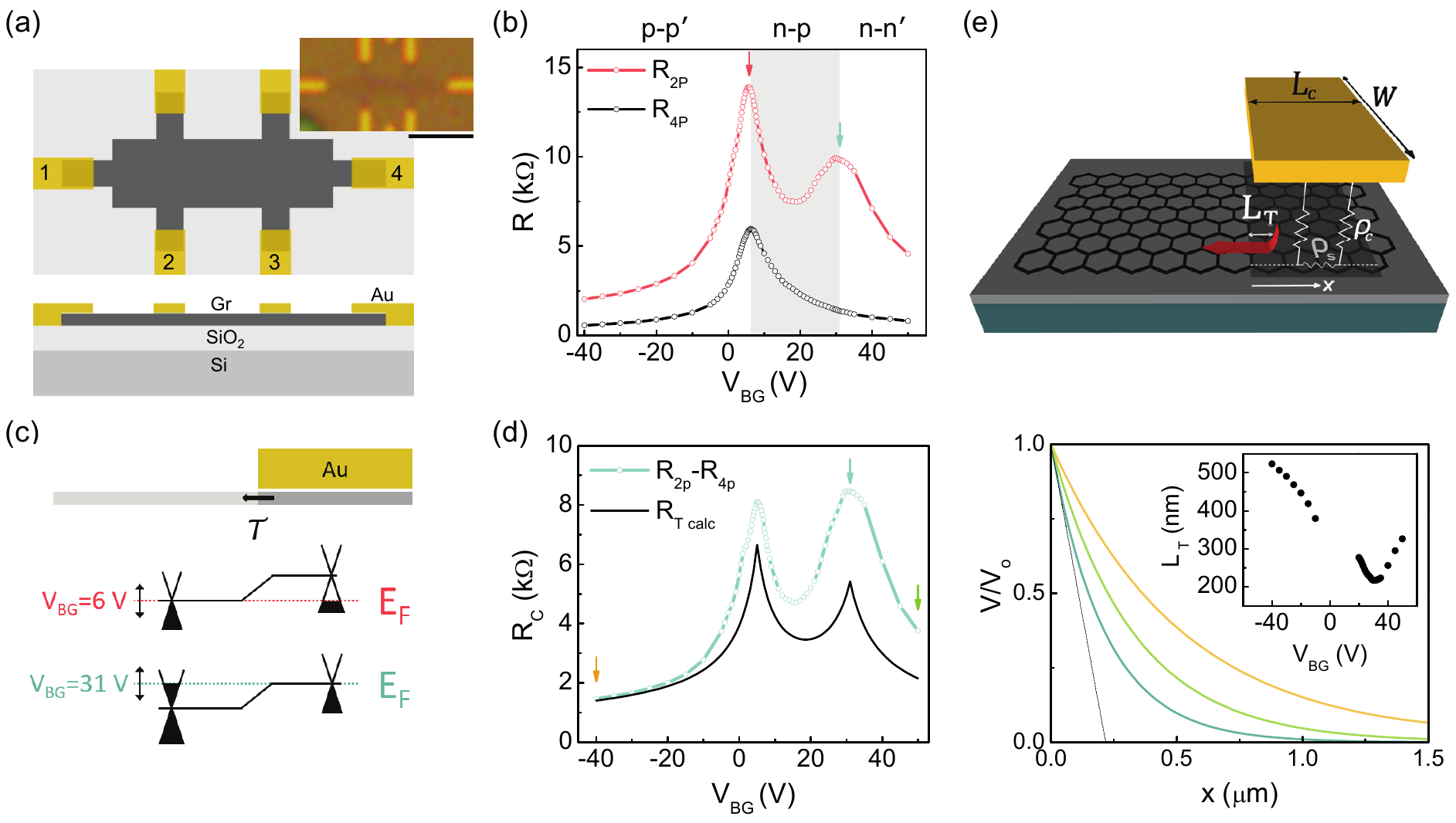}
\caption{Contact resistance and current crowding in Au-contacted graphene.
(a) Schematic of the device geometry and contact configuration. Inset
shows the optical image, scale bar is 5~$\mathrm{\lyxmathsym{\textmu}}$m.
(b) Resistance as a function of back gate voltage ($V_{{\rm BG}}$) measured
in 2-probe ($R_{{\rm 2P}}$) and the 4-probe ($R_{{\rm 4P}}$) geometry. (c) The
Fermi energy of graphene under the Au contacts can be tuned by applying
a back gate voltage. (d) The difference of the $R_{{\rm 2P}}$ and $R_{{\rm 4P}}$
gives the measured contact resistance $R_{{\rm c}}$ (blue). The resistance
$R_{{\rm T\,calc}}$ calculated for transport across the potential step
(black). (e) Current injection into graphene occurs within a
small length $\sim L_{{\rm T}}$ for a contact length of $L_{{\rm c}}$ (top). 
Most of the potential drop occurs at the edge of the contact, shown
at three $V_{{\rm BG}}$ values (bottom), marked in (d). Inset shows signification
variation in $L_{{\rm T}}$ with $V_{{\rm BG}}$ (Data near the main Dirac peak
excluded).}
\end{figure*}

\section*{Results}

\textbf{Characterization of Au-contacted Graphene.}
We first focus on a single layer graphene channel on conventional
(300~nm) SiO$_{2}/p^{++}$-Si substrate, etched into a Hall bar shape
with surface-contacted Au (99.999\%) leads (Fig.~1a). Here we used
pure gold contact (without a wetting underlayer of, \textit{e.g.}
Cr or Pd) because gold (hole) dopes the graphene underneath without
pinning the Fermi energy, or causing substantial modification in the
bandstructure~\cite{graphene-gold,Metal_doping_th}. This allows
easy tuning of the doping, and correspondingly the resistance, of
the contact region with backgate voltage ($V_{{\rm BG}}$). We measure the
two probe ($R_{{\rm 2P}}=R_{23,23}$) and four probe ($R_{{\rm 4P}}=R_{23,14}$)
resistance and noise as a function of $V_{{\rm BG}}$ between the leads
2 and 3 (suffixes in $R_{\rm{V^{+}V^{-},I^{+}I^{-}}}$ indicate the voltage
($V^{+},V^{-}$) and current ($I^{+},I^{-}$) leads). The $V_{{\rm BG}}$-dependence
of $R_{{\rm 2P}}$ and $R_{{\rm 4P}}$ are shown in Fig.~1b. $R_{{\rm 4P}}$ shows
a slightly asymmetric transfer characteristics, known to occur for
asymmetric contact doping~\cite{XiaNatNano2011}, with a single Dirac
point at $V_{{\rm BG}}\approx6$~V. $R_{{\rm 2P}}$, however, shows a second
Dirac point at $V_{{\rm BG}}\approx31$~V, due to the combination of hole
doping and weak pinning by Au at the contact region~\cite{graphene-gold,Metal_doping_th,XiaNatNano2011,SongCarbonLett2013},
which divides the transfer behavior in three parts ( $p-p'$, $n-p$
and $n-n'$), based on the sign of doping in the channel and contact
regions. The position of the Fermi level at the two Dirac points is
shown in the schematic of Fig.~1c. The observation of double Dirac
point in $R-V_{{\rm BG}}$ characteristics confirms the structural integrity
and gate-tunability of the Fermi level of the graphene channel underneath
the contact.

\begin{figure*}[t]
\includegraphics[width=2\columnwidth]{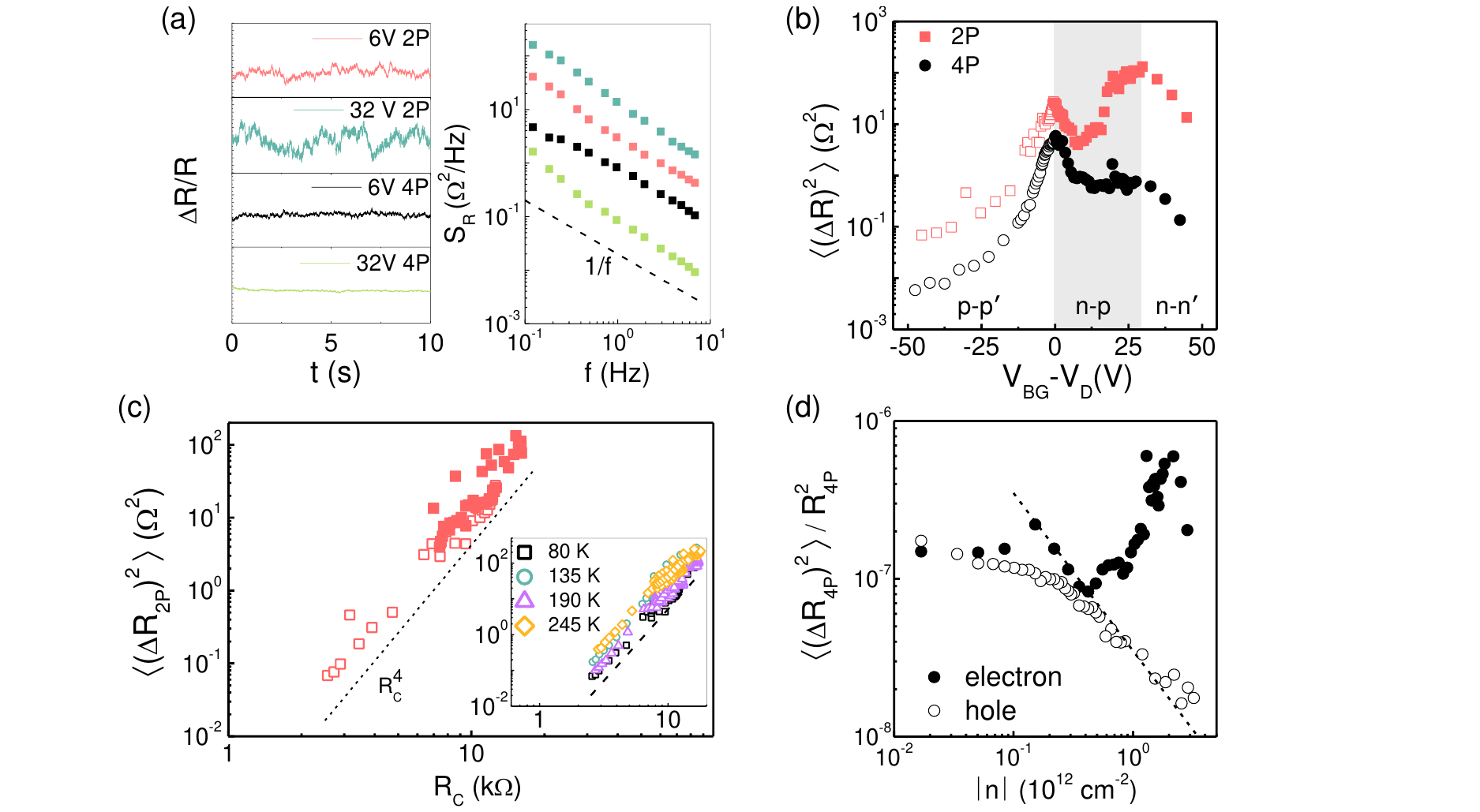}

\label{Figure2}

\caption{Contact noise in Au-contacted graphene.
(a) Typical measured time series (left) and corresponding power spectral
density $S_{{\rm R}}$ (right) as a function of frequency. (b) Noise (variance)
as a function of gate voltage in 2-probe $\langle(\Delta R_{{\rm 2P}})^{2}\rangle$
and 4-probe $\langle(\Delta R_{{\rm 4P}})^{2}\rangle$ geometry. (c)$\langle(\Delta R_{{\rm 2P}})^{2}\rangle$
as a function of contact resistance $R_{{\rm c}}$, shows $R_{{\rm c}}^{4}$ dependence.
This dependence is valid for temperatures down to $\sim80$~K (inset).
(d) Noise (4p) normalised by the graphene resistance as a function
of number density. }
\end{figure*}

\textbf{Contact resistance with Au-contacts and current crowding}
The shift in the local chemical potential by metal
contacts results in a change in resistance that contributes to contact
resistance, and determines the extent of current crowding at the gold-graphene
interface. To compute the contact resistance $R_{{\rm c}}$, we follow the
Landauer approach where the net transmission probability $\mathrm{T}$
across the contact is determined by the interplay of the number of
propagating modes in the channel and metal regions (Fig.~1c)~\cite{XiaNatNano2011}.
Fig.~1d shows the $V_{{\rm BG}}$-dependence of the experimental contact
resistance $R_{{\rm c}}=R_{{\rm 2P}}-R_{{\rm 4P}}$ (corrected for the resistance of
the small region of the probe arms) and that calculated assuming
the Dirac-like dispersion and level broadening $\approx80$~meV underneath
the contact and $\approx57$~meV in the channel (estimated from the
experimental transfer characteristics, see Supplementary Note~2
for the full details of calculations). The agreement, both in $V_{{\rm BG}}$-dependence
and absolute magnitude (within 50\% for all $V_{{\rm BG}}$), indicates
that the contact resistance is primarily composed of the resistance
$R_{{\rm T}}$ of the graphene layer over the charge transfer length ($L_{{\rm T}}$)
underneath the contacts. Due to mismatch between the resistivities
of the metal and graphene, $L_{{\rm T}}$ is significantly smaller than
the geometric width $L_{{\rm c}}$ ($\sim1-1.5~\mathrm{\lyxmathsym{\textmu}}$m)
of the metal lead, resulting in the current crowding effect.

To visualize this quantitatively, we consider the transmission line
model where the graphene layer below the contacts is represented with
a network of resistors characterized by sheet resistivity $\rho_{{\rm T}}$
(schematic in Fig.~1e). The potential profile in graphene under the
metal is then given by~\cite{current_crowd},

\begin{equation}
V(x)=\frac{\sqrt{\rho_{{\rm c}}R_{{\rm T}}}\text{cosh}((L_{{\rm c}}-x)/L_{{\rm T}})}{W\text{sinh}(L_{{\rm c}}/L_{{\rm T}})}I\label{voltage}
\end{equation}

\noindent where $I$ is the current flowing, $\rho_{{\rm c}}\approx200$~$\Omega\,\text{\textmu}$m$^{2}$~\cite{graphene-gold}
is the specific contact resistivity, and $L_{{\rm T}}=\sqrt{\rho_{{\rm c}}/R_{{\rm T}}}$
is charge transfer length from the edge by which $1/e$ of the current
is transferred to the metal contact ($W$ is the contact width). Taking
$R_{{\rm T}}$ as the experimentally observed contact resistance $R_{{\rm c}}$
(Fig.~1d), we calculated the potential drop underneath the contact,
normalized to its value at the edge $x=0$, for three gate voltages
marked by the arrows in Fig.~1d. The potential drops exponentially
over the gate voltage-dependent scale $L_{{\rm T}}$, being minimum ($\sim200$~nm)
for the second Dirac point at $+31$~V where the mismatch between
the resistivity of the metal and that of the graphene layer underneath
is maximum.

Since $R_{{\rm c}}=\sqrt{\rho_{{\rm c}}\rho_{{\rm T}}}/W$~\cite{current_crowd,Grosse_current_crowding}
and $\rho_{{\rm T}}=R_{{\rm T}}W/L_{{\rm T}}$, it is evident that the contact noise
is essentially the resistance fluctuations in the graphene layer underneath
the contact, \textit{i.e.} $\langle(\Delta R_{{\rm c}})^{2}\rangle/R_{{\rm c}}^{2}\approx\langle(\Delta R_{{\rm T}})^{2}\rangle/R_{{\rm T}}^{2}\propto\gamma_{{\rm T}}/n_{{\rm T}}$,
where $\gamma_{{\rm T}}$ and $n_{{\rm T}}$ are the phenomenological Hooge parameter
and carrier density in the charge transfer region, respectively. $\gamma_{{\rm T}}$
is independent of $n_{{\rm T}}$ and is determined by the kinetics of local
disorder induced by trapped charges, chemical modifications and changes
in the band dispersion due to hybridization. Assuming a diffusive
transport in the charge transfer region with density-independent mobility~\cite{Metal_doping_th},
the contact noise can be expressed as,

\begin{equation}
\frac{\langle(\Delta R_{{\rm c}})^{2}\rangle}{R_{{\rm c}}^{2}}\propto\frac{1}{n_{{\rm T}}}\propto\rho_{{\rm T}}\propto R_{{\rm c}}^{2}\label{contact}
\end{equation}

\noindent and implies a scaling relation $\langle(\Delta R_{{\rm c}})^{2}\rangle\propto R_{{\rm c}}^{4}$,
that can be readily verified experimentally. Note that: (1) The scaling
is different from that suggested for metal and 3D semiconductors where
the exponent of $R_{{\rm c}}$ is $\approx1$ for interface-type contacts
or $\approx3$ for constriction-type contacts~\cite{vandamme_contact_noise}.
(2) Since $n_{{\rm T}}$ is the only gate-tunable parameter~\cite{XiaNatNano2011,SongCarbonLett2013,WFpinning},
the scaling of contact resistance and electrical noise can be dynamically
monitored by varying the gate voltage, circumventing the necessity
to examine multiple pairs of contacts to isolate the contact contribution
to noise. (3) Although the absolute magnitude of the contact noise
is device/contact specific, the scaling of Eq.~(\ref{contact}) is
expected to hold irrespective of the geometry, material or chemical
nature of the contact (wetting or non-wetting).

\textbf{Noise measurement in Au-contacted graphene} Noise in
both $R_{{\rm 2P}}$ and $R_{{\rm 4P}}$ at all $V_{{\rm BG}}$ consists of random time
dependent fluctuations with power spectral density $S_{{\rm R}}(f)\propto1/f^{\alpha}$
(Fig.~2a), where $\alpha\approx1$ indicates usual $1/f$-noise due
to many independent fluctuators with wide distribution of characteristic
switching rates. However, to estimate and compare the total noise
magnitude, we have evaluated the ``variance'' $\langle(\Delta R)^{2}\rangle=\int{S_{{\rm R}}(f)df}$,
by integrating $S_{{\rm R}}(f)$ numerically over the experimental bandwidth.

Fig.~2b shows the $\Delta V_{{\rm BG}}$-dependence of $\langle(\Delta R_{{\rm 2P}})^{2}\rangle$
and $\langle(\Delta R_{{\rm 4P}})^{2}\rangle$, where the maxima in both
quantities align well with the Dirac points in $R_{{\rm 2P}}$ and $R_{{\rm c}}$
(Fig.~1b and e). The origin of the maximum in noise at the Dirac
point is a debated topic, and has often been attributed to low screening
ability of the graphene channel to fluctuating Coulomb potential at
the channel-substrate interface~\cite{M-shape_noise,KaverzinPRB2012,PalACSNano2011,RumyantsevJoP2010}.
Here, $\langle(\Delta R_{{\rm 2P}})^{2}\rangle$ peaks in the $n-p$ region
close to $\Delta V_{{\rm BG}}\approx25-30$~V, where the density of states
in the charge transfer region is low~\cite{XiaNatNano2011,SongCarbonLett2013},
indicating contact noise that originates due to poorly screened fluctuations
in the local Coulomb disorder. In fact, the noise magnitude at the
second peak ($\Delta V_{{\rm BG}}\approx30$~V) is $\sim10$ times larger
than that at the main Dirac peak, indicating the significantly larger
noise where the current crowding is most severe and contact resistance is the largest.
Surprisingly, $\langle(\Delta R_{{\rm 4P}})^{2}\rangle$ shows a weak increase
in this regime as well, suggesting a leakage
of the noise at the contacts even in four probe measurements (discussed
in more detail in the context of Fig.~3c and in the Supplementary
Note~3).

To verify the contact origin of noise, we have plotted $\langle(\Delta R_{{\rm 2P}})^{2}\rangle$
as a function of contact resistance $R_{{\rm c}}$ in Fig.~2c. Remarkably,
$\langle(\Delta R_{{\rm 2P}})^{2}\rangle$ for all $V_{{\rm BG}}$ collapses on
a single trace, and varies as $\langle(\Delta R_{{\rm 2P}})^{2}\rangle\propto R_{{\rm c}}^{4}$
over four decades of noise magnitude, suggesting that the measured
noise in two-probe configuration originates almost entirely
at the contacts, which is at least a factor of $10-100$ higher than
the channel noise $\langle(\Delta R_{{\rm 4P}})^{2}\rangle$ (circles in
Fig.~2b). Similar behavior was observed over a wide temperature range
as shown in the inset of Fig.~2c. In order to analyze the channel
contribution to noise, $\langle(\Delta R_{{\rm 4P}})^{2}\rangle$ is shown
as a function of the carrier density $n$ in Fig.~2d. For large hole
doping ($\gtrsim10^{12}$~cm$^{-2}$) \textit{i.e. p-p} regime where
$R_{{\rm c}}$ reduces to $\lesssim1$~k$\Omega$, we observed $\langle(\Delta R_{{\rm 4P}})^{2}\rangle/R_{{\rm 4P}}^{2}\propto1/n$
(dashed line), suggesting Hooge-type mobility fluctuation noise in
the graphene channel, with a Hooge parameter $\sim10^{-3}$~\cite{hooge,LinNanoLett2008,LiuAPL2009,ultralow_Atin,PalACSNano2011}.
However, in the \textit{n-n} regime, where the contact contribution
is dominant, noise deviates from $1/n$ behaviour.

\noindent 
\begin{figure*}[t]
\includegraphics[width=2\columnwidth]{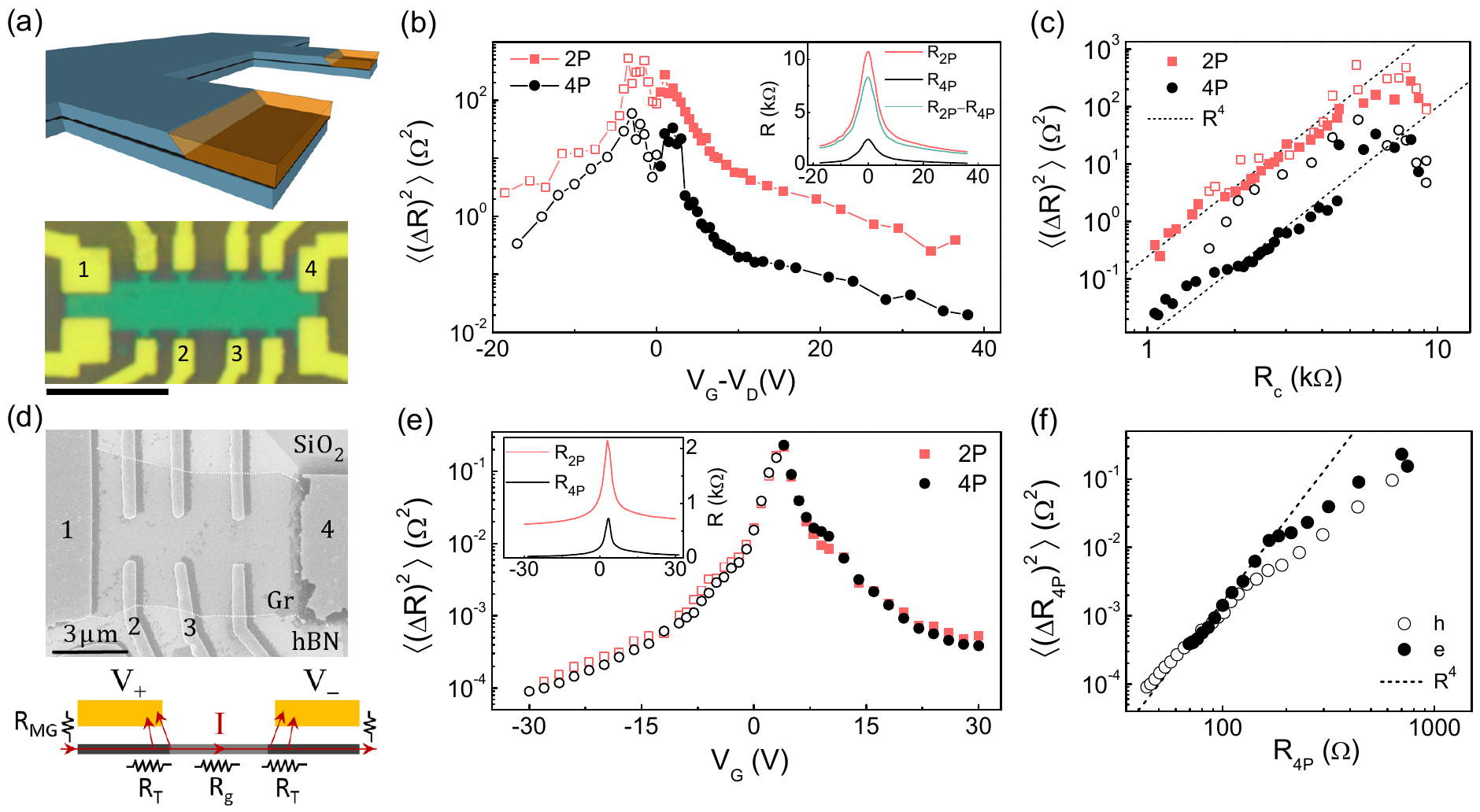}

\caption{Noise in high-mobility graphene (a) Hall bar device with graphene
encapsulated by hexagonal boron nitride (BN) and contacted by etching
the top BN, scale bar is 10~$\mathrm{\lyxmathsym{\textmu}}$m in
the optical image (bottom) (b) Noise variance $\langle(\Delta R)^{2}\rangle$
as a function of gate voltage in 2-probe and 4-probe geometry. Inset
shows resistance ($R_{{\rm 2P}}$, $R_{{\rm 4P}}$ and $R_{{\rm c}}$) as a function
of back gate voltage. (c) Noise variance $\langle(\Delta R)^{2}\rangle$
as a function of contact resistance $R_{{\rm c}}$, shows $R_{{\rm c}}^{4}$ dependence
(dotted line) for both two probe and four probe measurements. (d)
Electron microscope image of graphene on BN with invasive linear contacting
geometry. Bottom schematic shows the effect of charge scattering,
from the region under the contacts, on four probe measurements as
well. (e) $\langle(\Delta R)^{2}\rangle$ measured in 2-probe and
4-probe geometry roughly coincide for all gate voltages, indicating
noise generated in the channel is negligible. Inset shows $\text{R-V}_{\text{G}}$
characteristics for the device. (f) Noise shows $R^{4}$ dependence
(dotted line) at higher gate voltages indicating the dominant contact
contribution.}
\end{figure*}

\noindent 
\begin{figure*}[t]
\includegraphics[width=2\columnwidth]{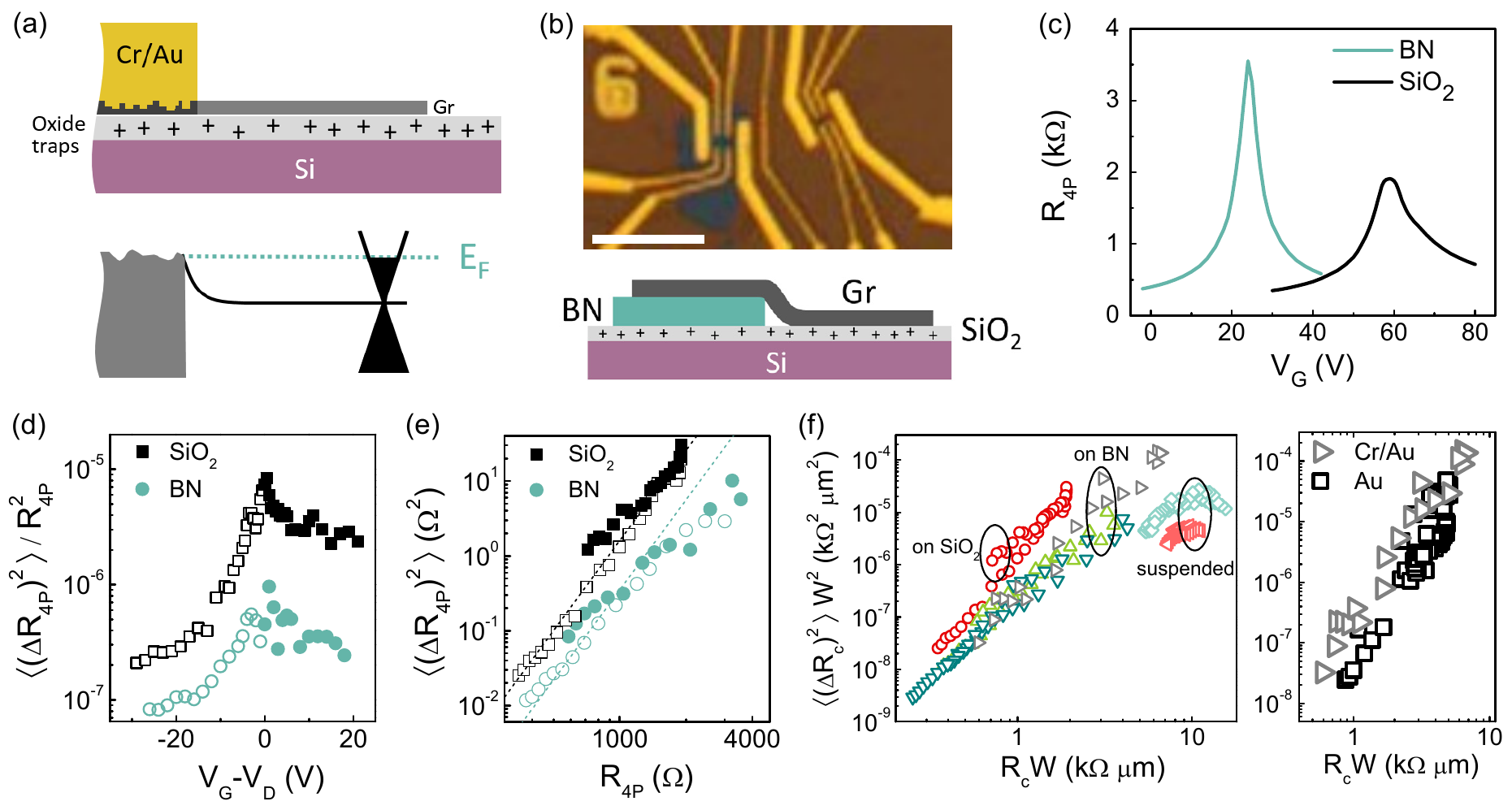}\caption{Noise mechanism (a) Schematic of the contact region with metal deposition,
and possible effects on the band dispersion and structural properties
of graphene underneath. (b) Device image and schematic showing a portion
of graphene on BN and the rest on SiO$_{2}$. Scale bar is 10~$\mathrm{\lyxmathsym{\textmu}}$m.
(c) Resistance as a function of gate voltage on BN (blue) and on SiO$_{2}$ (black).
(d) $\sim5$ to $\sim20$ times lower noise in graphene on BN (black
squares) than on SiO$_{2}$ (blue circles). (e) $\langle(\Delta R_{{\rm 4P}})^{2}\rangle$
as a function of $R{}_{{\rm 4P}}$ for graphene on SiO$_{2}$ (squares) and
for graphene on BN (circles), the dotted lines are $R^{4}$. (f) Comparison
of contact noise in various devices, noise depends strongly on the
substrate (left) or the contact material used (right).}
\end{figure*}

\textbf{Noise in high mobility graphene hybrids}
The dependence of the noise magnitude with the
contacting geometry in the Au-contacted device (Fig.~2) led us to
explore the contact contribution to noise in three other device geometries:
(1) Graphene, encapsulated between two hexagonal boron nitride (BN)
layers and etched into a Hall bar, contacted by etching only
the top BN (see Methods and Supplementary Note~1), shown in Fig.~3a.
(2) Graphene on SiO$_{2}$ and BN substrates (Fig.~3d and Fig.~4b),
in surface-contacted linear geometry, where the contacts extend on
to the channel region (invasive contacts), and (3) suspended graphene
devices which are intrinsically in two-probe contact configuration.
A 5~nm Cr underlayer was used with 50~nm Au films as contact material
in all these devices. The details of the device fabrication process
are given in the Methods section and Supplementary Note~1. Similar
to the Au contacted device, the noise measurements were performed
from 80~K to room temperature and no appreciable qualitative difference
was observed.

To examine the generality of the $R_{{\rm c}}^{4}$ scaling in high mobility
graphene FETs, we first measured both two probe and four probe noise
in the BN-encapsulated graphene hall bar device, which exhibited room
temperature (four probe) carrier mobilities of 58,000~cm$^{2}$V$^{-1}$s$^{-1}$
and 35,000~cm$^{2}$V$^{-1}$s$^{-1}$ in the electron doped and hole doped
regimes, respectively. The transfer characteristics shows only one
Dirac point ($V_{{\rm D}}$) for both $R_{{\rm 2P}}$ and $R_{{\rm 4P}}$ (Fig.~3b
inset), as expected for a Cr underlayer~\cite{WFpinning}. Both $\langle(\Delta R_{{\rm 2P}})^{2}\rangle$
and $\langle(\Delta R_{{\rm 4P}})^{2}\rangle$ decrease with increasing
$\lvert V_{{\rm G}}-V_{{\rm D}}\lvert$ (Fig.~3b), except over a small region
around $V_{{\rm D}}$ where the the distribution of charge in graphene becomes
inhomogeneous. Away from the inhomogeneous regime, both $\langle(\Delta R_{{\rm 2P}})^{2}\rangle$
and $\langle(\Delta R_{{\rm 4P}})^{2}\rangle$ exhibit the $R_{{\rm c}}^{4}$
scaling over three decades (Fig.~3c). The $R_{{\rm c}}^{4}$ scaling of
$\langle(\Delta R_{{\rm 4P}})^{2}\rangle$ is unexpected, although the suppression
$\langle(\Delta R_{{\rm 4P}})^{2}\rangle/\langle(\Delta R_{{\rm 2P}})^{2}\rangle\sim0.01$
is close to the nonlocal factor $\propto\exp[-2\pi L_{{\rm T}}/W]$ for
realistic $L_{{\rm T}}$ of $\sim400$~nm~\cite{Grosse_current_crowding,TransferLength_Pd_Ti},
suggesting this could be a nonlocal effect due to finite dimensions
of the voltage leads 2 and 3~\cite{VanderPauw} (see Supplementary
Note~3). This also explains the reduced, but perceptible
signature of contact noise in $\langle(\Delta R_{{\rm 4P}})^{2}\rangle$
in Fig.~2b and 2d. It is also interesting to note the drop in contact
noise magnitude in the inhomogeneous regime, which could be due to
the dominance of McWhorter-type number fluctuation noise~\cite{M-shape_noise,pellegrini_noise,KaverzinPRB2012,PalACSNano2011,bruno2},
rather than just mobility fluctuations in the charge transfer region.

The effect of contact noise becomes more severe for invasive surface
contacts (leads extending to the current flow path), as demonstrated
with a device that has graphene on BN (Fig.~3d). The transfer characteristics
shows a single Dirac point with carrier mobility $\sim35,000$~cm$^{2}$V$^{-1}$s$^{-1}$
(Fig.~3e inset). Strikingly, the magnitudes of $\langle(\Delta R_{{\rm 2P}})^{2}\rangle$
and $\langle(\Delta R_{{\rm 4P}})^{2}\rangle$ were found to be almost equal
over the entire range of $V_{{\rm BG}}$ (Fig.~3e), suggesting that the
dominant contribution to noise arises from the charge transfer region
underneath leads 2 and 3. To establish this quantitatively, we note
that $R_{{\rm 2P}}\approx2R_{{\rm MG}}+2R_{{\rm T}}+R_{{\rm g}}$ and $R_{{\rm 4P}}\approx2R_{{\rm T}}+R_{{\rm g}}$,
respectively (see schematics in Fig.~3d), where $R_{{\rm MG}}$ ($\lesssim300~\Omega$)
and $R_{{\rm g}}$ are the metal-graphene interface resistance and graphene
channel resistance, respectively. Due to the inseparability of $R_{{\rm T}}(=R_{{\rm c}})$
and $R_{{\rm g}}$ within this contacting scheme, we plot $\langle(\Delta R_{{\rm 4P}})^{2}\rangle$
as a function of $R_{{\rm 4P}}$ in Fig.~3f. It is evident that $\langle(\Delta R_{{\rm 4P}})^{2}\rangle\propto R_{{\rm 4P}}^{4}$
for $R_{{\rm 4P}}\lesssim150-200~\Omega$, where $R_{{\rm g}}$ is small due to
heavy electrostatic doping of the channel. However, for $R_{{\rm 4P}}\gtrsim200~\Omega$,
the deviation from the $R_{{\rm 4P}}^{4}$ scaling is likely due to finite
$R_{{\rm g}}$ that causes $R_{{\rm 4P}}$ to overestimate the true $R_{{\rm c}}$.
We have observed an $R^{4}$ scaling of noise for high-mobility suspended
graphene devices as well (see Fig.~4f).

\section*{Discussion}

Contact noise at the metal-semiconductor interface has been extensively
researched over nearly seven decades~\cite{hooge1969contact_noise,macfarlane1950contact_noise_theory,vandamme_contact_noise,vandamme1974noise_pc,luo1988theory,Bozhkov_noise_MS_contacts,guttler_noise_Schottky,TersoffNanoLett2007,ghatak_noise_MOS2},
and except for a few early models based on kinetics of interface disorder
such as adsorbate atoms~\cite{macfarlane1950contact_noise_theory},
the most common mechanism is based on time-dependent fluctuations
in the characteristics of the Schottky barrier at metal-semiconductor
junctions~\cite{guttler_noise_Schottky,luo1988theory,Bozhkov_noise_MS_contacts,macfarlane1950contact_noise_theory}.
The linearity of $I-V$ characteristics (not shown) and temperature
independence of $R_{{\rm c}}$ (see Supplementary Figure~4) in our devices
however eliminate the possibility of Schottky barrier-limited transport.
An alternative source of time varying potential is the trapped charge
at the SiO$_{2}$ surface~\cite{jayaraman19891,hung1990unifiedMOSnoise,Bosman_simulation,JDeen_MOS2,paul2015percolative,low_noise_SiP},
which has been suggested to cause contact noise even in ballistic
semiconducting carbon nanotubes FETs~\cite{TersoffNanoLett2007,Tersoff_Avouris_CNT}.
The reaction of graphene with metals spontaneously leads to chemical
modification (\textit{e.g.} carbide formation) and introduction of
defects (see schematic in Fig.~4a). The chemical modification and
defect formation can strongly influence the band structure of graphene
underneath the metal, suppressing the screening of Coulomb impurities.
This makes the charge transfer region susceptible to mobility fluctuations
due to trapped charge fluctuations in SiO$_{2}$, as indeed shown
recently for noise at grain boundaries in graphene~\cite{vidya_GBnoise}.

To verify this we have fabricated an invasively Cr/Au contacted device
where a single graphene channel was placed partially on BN (thickness
$\sim10$~nm), thus physically separating the channel from the oxide
traps~\cite{deanBN}, whereas the other part was directly in contact
with SiO$_{2}$ (Fig.~4b). The four probe transfer characteristics
(Fig.~4c) confirms that the region of graphene placed on SiO$_{2}$
shows lower carrier mobility (7500~cm$^{2}$V$^{-1}$s$^{-1}$ and 4000~cm$^{2}$V$^{-1}$s$^{-1}$
for hole and electron doping respectively) than the corresponding
mobility (8000~cm$^{2}$V$^{-1}$s$^{-1}$ and 7500~cm$^{2}$V$^{-1}$s$^{-1}$)
of the part on BN, as well as strong substrate-induced doping, both
of which can be readily understood by the proximity to charge traps
at the SiO$_{2}$ surface. Although $\langle(\Delta R_{{\rm 4P}})^{2}\rangle$
in both parts show strong peaks at the respective Dirac points (Fig.~4d),
it is evident that the normalized noise magnitude in the graphene
on SiO$_{2}$ substrate is up to a factor of ten larger than
that on BN, similar to that reported recently~\cite{edge_contact_noise,NoiseGBN}.
The scaling $\langle(\Delta R_{{\rm 4P}})^{2}\rangle\propto R_{{\rm 4P}}^{4}$
(Fig.~4e) over three decades of noise magnitude, irrespective of
the substrate, unambiguously indicates the dominance of contact noise,
and that the contact noise in graphene FETs is primarily a result
of mobility fluctuations in the charge transfer region due to fluctuating
Coulomb potential from local charge traps (predominantly from the
SiO$_{2}$ surface).

Finally, in order to outline a recipe to minimize the contact noise
in graphene devices, we have compiled the normalised magnitude of
specific contact noise $\langle(\Delta R_{{\rm c}})^{2}\rangle W^{2}$ as
a function of specific contact resistance $R_{{\rm c}}W$, from different
classes of devices that were studied in this work. We identify two
key factors that impact the contact noise: Firstly, as can be clearly
seen in Fig.~4f (left), the specific contact noise is largest for graphene
on SiO$_{2}$, lower on devices with graphene on BN and lowest for
suspended graphene devices where all SiO$_{2}$ has been etched away
from under the graphene channel as well as partially from below the
contact region (see Supplementary Figure~5). Moreover, noise data
from all devices with BN as substrate collapse on top of each other,
regardless of mobility values, indicating that the separation of contacts
from the SiO$_{2}$ traps is the primary factor that determines the
noise magnitude rather than the channel quality itself. Secondly,
it can also be seen from Fig.~4f (right), that the device with
Cr/Au contacts, which are known to chemically modify graphene~\cite{WFpinning,GongAcsNano2014},
exhibits higher noise than the device with Au contacts, which is expected
to leave graphene intact, despite the fact that the former device
has a BN substrate whereas the later SiO$_{2}$. This highlights the
major role of defects under metal contacts in noise generation. Combining
these factors leads to the conclusion that minimizing environmental
electrostatic fluctuations and developing a contacting scheme that
preserves the chemical/structural integrity of graphene, will be necessary
for ultra-low noise graphene electronics.

In conclusion, we have studied electrical noise at the metal contacts
in graphene devices with a large range of carrier mobility, on multiple
substrates with various device and lead geometries. Using a phenomenological
model of contact noise for purely two dimensional materials we show
that contact noise is often the dominant noise source in graphene
devices. The influence of contact noise is most severe in high-mobility
graphene transistors. Most surprisingly, we discover the ubiquity
of contact noise which is seen to affect even four probe measurements
in a Hall bar geometry. Our analysis suggests that contact noise is
caused by strong mobility fluctuations in the charge transfer region
under the metal contacts, due to the fluctuating electrostatic environment.
A microscopic understanding of contact noise may aid in the development
of ultra-low noise graphene electronics.

\section*{Methods}

\paragraph{\textbf{\textup{Device Fabrication.}} \textup{Graphene and hexagonal
boron nitride (BN) were exfoliated on SiO$_{2}$ using the 3M scotch
(Magic) tape. The heterostructures were assembled using a method similar
to that described in ref.~\cite{ZomerAPL2011} in a custom built
microscope and transfer assembly. For parameters similar to those
described in ref.~\cite{graphene_edge_contact} we determined the
etching rate of BN, in a CHF$_{3}$ and O$_{2}$ plasma, to be $23\pm2$~nm
per 60s (see Supplementary Figure~1). The device shown in Fig.~3a was fabricated
by etching only the top BN ($21\pm3$~nm, etched for 60s). Two layers
of PMMA (450 K and 950 K) were spin coated for electron beam lithography
and act as masks for metal deposition and etching. Graphene was contacted
by thermally evaporating Au (50~nm) or Cr/Au (5/50~nm) at $\lesssim10^{-6}$~mbar.}}

\paragraph{\textbf{\textup{Measurements.}} \textup{Both average resistance and
time-dependent noise were measured in standard low-frequency lock-in
technique, with a small source-drain excitation current $\sim100$~nA
to ensure linear transport regime~\cite{ArindamArxiv}. Background
noise was measured simultaneously and was subtracted from total noise
to determine the sample noise.}\protect \\
}

\paragraph{\textbf{\textup{Data availability.}} \textup{The data that support
the findings of this study are available from the corresponding author upon request.}\protect \\}

\begin{acknowledgments}
We thank the Department of Science and Technology and Tokyo Electron
Limited for financial support. S.G. thanks the IISc Centenary Postdoctoral
Fellowship for financial support. S.Gh. thanks CSIR for financial
support. 
\end{acknowledgments}

\section*{Contributions}

P.K., T.P.S., S.G., S.Gh., S.K. and A.G. designed the experiments.
P.K., T.P.S., S.G. and S.Gh. fabricated the devices and performed
the measurements. P.K., T.P.S., S.G., S.Gh. and A.G. analysed the
data and discussed the results. P.K., T.P.S., S.G. and A.G. wrote
the manuscript with inputs from all authors. P.K., T.P.S and S.G. contributed equally to this work.

\section*{Competing financial interests}
The authors declare no conflict of interests.


\begin{thebibliography}{10}
\expandafter\ifx\csname url\endcsname\relax
  \def\url#1{\texttt{#1}}\fi
\expandafter\ifx\csname urlprefix\endcsname\relax\def\urlprefix{URL }\fi
\providecommand{\bibinfo}[2]{#2}
\providecommand{\eprint}[2][]{\url{#2}}

\bibitem{Goddard_end_contact}
\bibinfo{author}{Matsuda, Y.}, \bibinfo{author}{Deng, W.-Q.} \&
  \bibinfo{author}{Goddard~III, W.~A.}
\newblock \bibinfo{title}{Contact resistance for "end-contacted" metal-graphene
  and metal-nanotube interfaces from quantum mechanics}.
\newblock \emph{\bibinfo{journal}{J. Phys. Chem. C}}
  \textbf{\bibinfo{volume}{114}}, \bibinfo{pages}{17845--17850}
  (\bibinfo{year}{2010}).

\bibitem{SmithAcsNano2013}
\bibinfo{author}{Smith, J.~T.}, \bibinfo{author}{Franklin, A.~D.},
  \bibinfo{author}{Farmer, D.~B.} \& \bibinfo{author}{Dimitrakopoulos, C.~D.}
\newblock \bibinfo{title}{Reducing contact resistance in graphene devices
  through contact area patterning}.
\newblock \emph{\bibinfo{journal}{ACS Nano}} \textbf{\bibinfo{volume}{7}},
  \bibinfo{pages}{3661--3667} (\bibinfo{year}{2013}).

\bibitem{graphene_edge_contact}
\bibinfo{author}{Wang, L.} \emph{et~al.}
\newblock \bibinfo{title}{One-dimensional electrical contact to a
  two-dimensional material}.
\newblock \emph{\bibinfo{journal}{Science}} \textbf{\bibinfo{volume}{342}},
  \bibinfo{pages}{614--617} (\bibinfo{year}{2013}).

\bibitem{edge_contact2}
\bibinfo{author}{Song, S.~M.}, \bibinfo{author}{Kim, T.~Y.},
  \bibinfo{author}{Sul, O.~J.}, \bibinfo{author}{Shin, W.~C.} \&
  \bibinfo{author}{Cho, B.~J.}
\newblock \bibinfo{title}{Improvement of graphene--metal contact resistance by
  introducing edge contacts at graphene under metal}.
\newblock \emph{\bibinfo{journal}{Appl. Phys. Lett.}}
  \textbf{\bibinfo{volume}{104}}, \bibinfo{pages}{183506}
  (\bibinfo{year}{2014}).

\bibitem{phase_noise}
\bibinfo{author}{Razavi, B.}
\newblock \bibinfo{title}{A study of phase noise in CMOS oscillators}.
\newblock \emph{\bibinfo{journal}{IEEE J. Solid-State Circuits}}
  \textbf{\bibinfo{volume}{31}}, \bibinfo{pages}{331--343}
  (\bibinfo{year}{1996}).

\bibitem{TersoffNanoLett2007}
\bibinfo{author}{Tersoff, J.}
\newblock \bibinfo{title}{Low-frequency noise in nanoscale ballistic
  transistors}.
\newblock \emph{\bibinfo{journal}{Nano Lett.}} \textbf{\bibinfo{volume}{7}},
  \bibinfo{pages}{194--198} (\bibinfo{year}{2007}).

\bibitem{RumyantsevJoP2010}
\bibinfo{author}{Rumyantsev, S.}, \bibinfo{author}{Liu, G.},
  \bibinfo{author}{Stillman, W.}, \bibinfo{author}{Shur, M.} \&
  \bibinfo{author}{Balandin, A.~A.}
\newblock \bibinfo{title}{Electrical and noise characteristics of graphene
  field-effect transistors: ambient effects, noise sources and physical
  mechanisms}.
\newblock \emph{\bibinfo{journal}{J. Phys. Conden. Matter}}
  \textbf{\bibinfo{volume}{22}}, \bibinfo{pages}{395302}
  (\bibinfo{year}{2010}).

\bibitem{PalACSNano2011}
\bibinfo{author}{Pal, A.~N.} \emph{et~al.}
\newblock \bibinfo{title}{Microscopic mechanism of 1/$f$ noise in graphene:
  Role of energy band dispersion}.
\newblock \emph{\bibinfo{journal}{ACS Nano}} \textbf{\bibinfo{volume}{5}},
  \bibinfo{pages}{2075--2081} (\bibinfo{year}{2011}).

\bibitem{HellerNanoLett2010}
\bibinfo{author}{Heller, I.} \emph{et~al.}
\newblock \bibinfo{title}{Charge noise in graphene transistors}.
\newblock \emph{\bibinfo{journal}{Nano Lett.}} \textbf{\bibinfo{volume}{10}},
  \bibinfo{pages}{1563--1567} (\bibinfo{year}{2010}).

\bibitem{graded_graphene}
\bibinfo{author}{Liu, G.}, \bibinfo{author}{Rumyantsev, S.},
  \bibinfo{author}{Shur, M.} \& \bibinfo{author}{Balandin, A.~A.}
\newblock \bibinfo{title}{Graphene thickness-graded transistors with reduced
  electronic noise}.
\newblock \emph{\bibinfo{journal}{Appl. Phys. Lett.}}
  \textbf{\bibinfo{volume}{100}}, \bibinfo{pages}{033103}
  (\bibinfo{year}{2012}).

\bibitem{LiuAPL2009}
\bibinfo{author}{Liu, G.} \emph{et~al.}
\newblock \bibinfo{title}{Low-frequency electronic noise in the double-gate
  single-layer graphene transistors}.
\newblock \emph{\bibinfo{journal}{Appl. Phys. Lett.}}
  \textbf{\bibinfo{volume}{95}}, \bibinfo{pages}{033103}
  (\bibinfo{year}{2009}).

\bibitem{Suspended_contact_noise}
\bibinfo{author}{Kumar, M.}, \bibinfo{author}{Laitinen, A.},
  \bibinfo{author}{Cox, D.} \& \bibinfo{author}{Hakonen, P.~J.}
\newblock \bibinfo{title}{Ultra low 1/$f$ noise in suspended bilayer graphene}.
\newblock \emph{\bibinfo{journal}{Appl. Phys. Lett.}}
  \textbf{\bibinfo{volume}{106}}, \bibinfo{pages}{263505}
  (\bibinfo{year}{2015}).

\bibitem{LinNanoLett2008}
\bibinfo{author}{Lin, Y.-M.} \& \bibinfo{author}{Avouris, P.}
\newblock \bibinfo{title}{Strong suppression of electrical noise in bilayer
  graphene nanodevices}.
\newblock \emph{\bibinfo{journal}{Nano Lett.}} \textbf{\bibinfo{volume}{8}},
  \bibinfo{pages}{2119--2125} (\bibinfo{year}{2008}).

\bibitem{PalPRL2009}
\bibinfo{author}{Pal, A.~N.} \& \bibinfo{author}{Ghosh, A.}
\newblock \bibinfo{title}{Resistance noise in electrically biased bilayer
  graphene}.
\newblock \emph{\bibinfo{journal}{Phys. Rev. Lett.}}
  \textbf{\bibinfo{volume}{102}}, \bibinfo{pages}{126805}
  (\bibinfo{year}{2009}).

\bibitem{M-shape_noise}
\bibinfo{author}{Xu, G.} \emph{et~al.}
\newblock \bibinfo{title}{Effect of spatial charge inhomogeneity on 1/$f$ noise
  behavior in graphene}.
\newblock \emph{\bibinfo{journal}{Nano Lett.}} \textbf{\bibinfo{volume}{10}},
  \bibinfo{pages}{3312--3317} (\bibinfo{year}{2010}).

\bibitem{KaverzinPRB2012}
\bibinfo{author}{Kaverzin, A.}, \bibinfo{author}{Mayorov, A.},
  \bibinfo{author}{Shytov, A.} \& \bibinfo{author}{Horsell, D.}
\newblock \bibinfo{title}{Impurities as a source of 1/$f$ noise in graphene}.
\newblock \emph{\bibinfo{journal}{Phys. Rev. B}} \textbf{\bibinfo{volume}{85}},
  \bibinfo{pages}{075435} (\bibinfo{year}{2012}).

\bibitem{BalandinNatNano2013}
\bibinfo{author}{Balandin, A.~A.}
\newblock \bibinfo{title}{Low-frequency 1/$f$ noise in graphene devices}.
\newblock \emph{\bibinfo{journal}{Nat. Nanotechnol.}}
  \textbf{\bibinfo{volume}{8}}, \bibinfo{pages}{549--555}
  (\bibinfo{year}{2013}).

\bibitem{pellegrini_noise}
\bibinfo{author}{Pellegrini, B.}
\newblock \bibinfo{title}{1/$f$ noise in graphene}.
\newblock \emph{\bibinfo{journal}{Eur. Phys. J. B}}
  \textbf{\bibinfo{volume}{86}}, \bibinfo{pages}{1--12} (\bibinfo{year}{2013}).

\bibitem{edge_contact_noise}
\bibinfo{author}{Stolyarov, M.~A.}, \bibinfo{author}{Liu, G.},
  \bibinfo{author}{Rumyantsev, S.~L.}, \bibinfo{author}{Shur, M.} \&
  \bibinfo{author}{Balandin, A.~A.}
\newblock \bibinfo{title}{Suppression of 1/$f$ noise in near-ballistic
  h-BN-graphene-h-BN heterostructure field-effect transistors}.
\newblock \emph{\bibinfo{journal}{Appl. Phys. Lett.}}
  \textbf{\bibinfo{volume}{107}}, \bibinfo{pages}{023106}
  (\bibinfo{year}{2015}).

\bibitem{NoiseGBN}
\bibinfo{author}{Kayyalha, M.} \& \bibinfo{author}{Chen, Y.~P.}
\newblock \bibinfo{title}{Observation of reduced 1/$f$ noise in graphene field
  effect transistors on boron nitride substrates}.
\newblock \emph{\bibinfo{journal}{Appl. Phys. Lett.}}
  \textbf{\bibinfo{volume}{107}}, \bibinfo{pages}{113101}
  (\bibinfo{year}{2015}).

\bibitem{ChandanNL}
\bibinfo{author}{Kumar, C.}, \bibinfo{author}{Kuiri, M.},
  \bibinfo{author}{Jung, J.}, \bibinfo{author}{Das, T.} \&
  \bibinfo{author}{Das, A.}
\newblock \bibinfo{title}{Tunability of 1/$f$ noise at multiple Dirac cones in
  hBN encapsulated graphene devices}.
\newblock \emph{\bibinfo{journal}{Nano Lett.}} \textbf{\bibinfo{volume}{16}},
  \bibinfo{pages}{1042--1049} (\bibinfo{year}{2016}).

\bibitem{current_crowd}
\bibinfo{author}{Murrmann, H.} \& \bibinfo{author}{Widmann, D.}
\newblock \bibinfo{title}{Current crowding on metal contacts to planar
  devices}.
\newblock \emph{\bibinfo{journal}{IEEE Trans. Electron Devices}}
  \textbf{\bibinfo{volume}{16}}, \bibinfo{pages}{1022--1024}
  (\bibinfo{year}{1969}).

\bibitem{vandammecurrentcrowd}
\bibinfo{author}{Vandamme, E.} \& \bibinfo{author}{Vandamme, L.}
\newblock \bibinfo{title}{Current crowding and its effect on 1/$f$ noise and
  third harmonic distortion--a case study for quality assessment of resistors}.
\newblock \emph{\bibinfo{journal}{Microelectron. Reliab.}}
  \textbf{\bibinfo{volume}{40}}, \bibinfo{pages}{1847--1853}
  (\bibinfo{year}{2000}).

\bibitem{Grosse_current_crowding}
\bibinfo{author}{Grosse, K.~L.}, \bibinfo{author}{Bae, M.-H.},
  \bibinfo{author}{Lian, F.}, \bibinfo{author}{Pop, E.} \&
  \bibinfo{author}{King, W.~P.}
\newblock \bibinfo{title}{Nanoscale joule heating, peltier cooling and current
  crowding at graphene-metal contacts}.
\newblock \emph{\bibinfo{journal}{Nat. Nanotechnol.}}
  \textbf{\bibinfo{volume}{6}}, \bibinfo{pages}{287--290}
  (\bibinfo{year}{2011}).

\bibitem{NagashioAPL2010}
\bibinfo{author}{Nagashio, K.}, \bibinfo{author}{Nishimura, T.},
  \bibinfo{author}{Kita, K.} \& \bibinfo{author}{Toriumi, A.}
\newblock \bibinfo{title}{Contact resistivity and current flow path at
  metal/graphene contact}.
\newblock \emph{\bibinfo{journal}{Appl. Phys. Lett.}}
  \textbf{\bibinfo{volume}{97}}, \bibinfo{pages}{143514}
  (\bibinfo{year}{2010}).

\bibitem{P_current_crowding}
\bibinfo{author}{Wang, Q.}, \bibinfo{author}{Tao, X.}, \bibinfo{author}{Yang,
  L.} \& \bibinfo{author}{Gu, Y.}
\newblock \bibinfo{title}{Current crowding in two-dimensional black-phosphorus
  field-effect transistors}.
\newblock \emph{\bibinfo{journal}{Appl. Phys. Lett.}}
  \textbf{\bibinfo{volume}{108}}, \bibinfo{pages}{103109}
  (\bibinfo{year}{2016}).

\bibitem{MoS2_current_crowding}
\bibinfo{author}{Yuan, H.} \emph{et~al.}
\newblock \bibinfo{title}{Field effects of current crowding in metal-{MoS}$_2$
  contacts}.
\newblock \emph{\bibinfo{journal}{Appl. Phys. Lett.}}
  \textbf{\bibinfo{volume}{108}}, \bibinfo{pages}{103505}
  (\bibinfo{year}{2016}).

\bibitem{berger1972contact}
\bibinfo{author}{Berger, H.}
\newblock \bibinfo{title}{Contact resistance and contact resistivity}.
\newblock \emph{\bibinfo{journal}{J. Electrochem. Soc.}}
  \textbf{\bibinfo{volume}{119}}, \bibinfo{pages}{507--514}
  (\bibinfo{year}{1972}).

\bibitem{Photo_contact}
\bibinfo{author}{Lee, E.~J.}, \bibinfo{author}{Balasubramanian, K.},
  \bibinfo{author}{Weitz, R.~T.}, \bibinfo{author}{Burghard, M.} \&
  \bibinfo{author}{Kern, K.}
\newblock \bibinfo{title}{Contact and edge effects in graphene devices}.
\newblock \emph{\bibinfo{journal}{Nat. Nanotechnol.}}
  \textbf{\bibinfo{volume}{3}}, \bibinfo{pages}{486--490}
  (\bibinfo{year}{2008}).

\bibitem{photocurrentxia2009}
\bibinfo{author}{Xia, F.} \emph{et~al.}
\newblock \bibinfo{title}{Photocurrent imaging and efficient photon detection
  in a graphene transistor}.
\newblock \emph{\bibinfo{journal}{Nano Lett.}} \textbf{\bibinfo{volume}{9}},
  \bibinfo{pages}{1039--1044} (\bibinfo{year}{2009}).

\bibitem{photoscanning}
\bibinfo{author}{Mueller, T.} \emph{et~al.}
\newblock \bibinfo{title}{Role of contacts in graphene transistors: A scanning
  photocurrent study}.
\newblock \emph{\bibinfo{journal}{Phys. Rev. B}} \textbf{\bibinfo{volume}{79}},
  \bibinfo{pages}{245430} (\bibinfo{year}{2009}).

\bibitem{Kelvin_probe}
\bibinfo{author}{Yu, Y.-J.} \emph{et~al.}
\newblock \bibinfo{title}{Tuning the graphene work function by electric field
  effect}.
\newblock \emph{\bibinfo{journal}{Nano Lett.}} \textbf{\bibinfo{volume}{9}},
  \bibinfo{pages}{3430--3434} (\bibinfo{year}{2009}).

\bibitem{Kelvin_probe1}
\bibinfo{author}{Yan, L.}, \bibinfo{author}{Punckt, C.},
  \bibinfo{author}{Aksay, I.~A.}, \bibinfo{author}{Mertin, W.} \&
  \bibinfo{author}{Bacher, G.}
\newblock \bibinfo{title}{Local voltage drop in a single functionalized
  graphene sheet characterized by Kelvin probe force microscopy}.
\newblock \emph{\bibinfo{journal}{Nano Lett.}} \textbf{\bibinfo{volume}{11}},
  \bibinfo{pages}{3543--3549} (\bibinfo{year}{2011}).

\bibitem{SongCarbonLett2013}
\bibinfo{author}{Song, S.~M.} \& \bibinfo{author}{Cho, B.~J.}
\newblock \bibinfo{title}{Contact resistance in graphene channel transistors}.
\newblock \emph{\bibinfo{journal}{Carbon Lett.}} \textbf{\bibinfo{volume}{14}},
  \bibinfo{pages}{162--170} (\bibinfo{year}{2013}).

\bibitem{vandamme_contact_noise}
\bibinfo{author}{Vandamme, L.}
\newblock \bibinfo{title}{Noise as a diagnostic tool for quality and
  reliability of electronic devices}.
\newblock \emph{\bibinfo{journal}{IEEE Trans. Electron Devices}}
  \textbf{\bibinfo{volume}{41}}, \bibinfo{pages}{2176--2187}
  (\bibinfo{year}{1994}).

\bibitem{GongAcsNano2014}
\bibinfo{author}{Gong, C.} \emph{et~al.}
\newblock \bibinfo{title}{Realistic metal-graphene contact structures}.
\newblock \emph{\bibinfo{journal}{ACS Nano}} \textbf{\bibinfo{volume}{8}},
  \bibinfo{pages}{642--649} (\bibinfo{year}{2014}).

\bibitem{HuardPRB2008}
\bibinfo{author}{Huard, B.}, \bibinfo{author}{Stander, N.},
  \bibinfo{author}{Sulpizio, J.~A.} \& \bibinfo{author}{Goldhaber-Gordon, D.}
\newblock \bibinfo{title}{Evidence of the role of contacts on the observed
  electron-hole asymmetry in graphene}.
\newblock \emph{\bibinfo{journal}{Phys. Rev. B}} \textbf{\bibinfo{volume}{78}},
  \bibinfo{pages}{121402} (\bibinfo{year}{2008}).

\bibitem{XiaNatNano2011}
\bibinfo{author}{Xia, F.}, \bibinfo{author}{Perebeinos, V.},
  \bibinfo{author}{Lin, Y.-m.}, \bibinfo{author}{Wu, Y.} \&
  \bibinfo{author}{Avouris, P.}
\newblock \bibinfo{title}{The origins and limits of metal-graphene junction
  resistance}.
\newblock \emph{\bibinfo{journal}{Nat. Nanotechnol.}}
  \textbf{\bibinfo{volume}{6}}, \bibinfo{pages}{179--184}
  (\bibinfo{year}{2011}).

\bibitem{WFpinning}
\bibinfo{author}{Song, S.~M.}, \bibinfo{author}{Park, J.~K.},
  \bibinfo{author}{Sul, O.~J.} \& \bibinfo{author}{Cho, B.~J.}
\newblock \bibinfo{title}{Determination of work function of graphene under a
  metal electrode and its role in contact resistance}.
\newblock \emph{\bibinfo{journal}{Nano Lett.}} \textbf{\bibinfo{volume}{12}},
  \bibinfo{pages}{3887--3892} (\bibinfo{year}{2012}).

\bibitem{Rc_Mayank}
\bibinfo{author}{Ghatge, M.} \& \bibinfo{author}{Shrivastava, M.}
\newblock \bibinfo{title}{Physical insights on the ambiguous metal--graphene
  interface and proposal for improved contact resistance}.
\newblock \emph{\bibinfo{journal}{IEEE Trans. Electron Devices}}
  \textbf{\bibinfo{volume}{62}}, \bibinfo{pages}{4139--4147}
  (\bibinfo{year}{2015}).

\bibitem{Contact_resistance_2part}
\bibinfo{author}{Russo, S.}, \bibinfo{author}{Craciun, M.},
  \bibinfo{author}{Yamamoto, M.}, \bibinfo{author}{Morpurgo, A.} \&
  \bibinfo{author}{Tarucha, S.}
\newblock \bibinfo{title}{Contact resistance in graphene-based devices}.
\newblock \emph{\bibinfo{journal}{Phys. E}} \textbf{\bibinfo{volume}{42}},
  \bibinfo{pages}{677--679} (\bibinfo{year}{2010}).

\bibitem{independent_contactR}
\bibinfo{author}{Venugopal, A.}, \bibinfo{author}{Colombo, L.} \&
  \bibinfo{author}{Vogel, E.}
\newblock \bibinfo{title}{Contact resistance in few and multilayer graphene
  devices}.
\newblock \emph{\bibinfo{journal}{Appl. Phys. Lett.}}
  \textbf{\bibinfo{volume}{96}}, \bibinfo{pages}{013512}
  (\bibinfo{year}{2010}).

\bibitem{ghatak_noise_MOS2}
\bibinfo{author}{Ghatak, S.}, \bibinfo{author}{Mukherjee, S.},
  \bibinfo{author}{Jain, M.}, \bibinfo{author}{Sarma, D.~D.} \&
  \bibinfo{author}{Ghosh, A.}
\newblock \bibinfo{title}{Microscopic origin of low frequency noise in
  {MoS}$_2$ field-effect transistors}.
\newblock \emph{\bibinfo{journal}{APL Mater.}} \textbf{\bibinfo{volume}{2}},
  \bibinfo{pages}{092515} (\bibinfo{year}{2014}).

\bibitem{paul2015percolative}
\bibinfo{author}{Paul, T.}, \bibinfo{author}{Ghatak, S.} \&
  \bibinfo{author}{Ghosh, A.}
\newblock \bibinfo{title}{Percolative switching in transition metal
  dichalcogenide field-effect transistors at room temperature}.
\newblock \emph{\bibinfo{journal}{Nanotechnology}}
  \textbf{\bibinfo{volume}{27}}, \bibinfo{pages}{125706}
  (\bibinfo{year}{2016}).

\bibitem{graphene-gold}
\bibinfo{author}{Sundaram, R.~S.} \emph{et~al.}
\newblock \bibinfo{title}{The graphene--gold interface and its implications for
  nanoelectronics}.
\newblock \emph{\bibinfo{journal}{Nano Lett.}} \textbf{\bibinfo{volume}{11}},
  \bibinfo{pages}{3833--3837} (\bibinfo{year}{2011}).

\bibitem{Metal_doping_th}
\bibinfo{author}{Giovannetti, G.} \emph{et~al.}
\newblock \bibinfo{title}{Doping graphene with metal contacts}.
\newblock \emph{\bibinfo{journal}{Phys. Rev. Lett.}}
  \textbf{\bibinfo{volume}{101}}, \bibinfo{pages}{026803}
  (\bibinfo{year}{2008}).

\bibitem{hooge}
\bibinfo{author}{Hooge, F.~N.}
\newblock \bibinfo{title}{1/$f$ noise is no surface effect}.
\newblock \emph{\bibinfo{journal}{Physics letters A}}
  \textbf{\bibinfo{volume}{29}}, \bibinfo{pages}{139--140}
  (\bibinfo{year}{1969}).

\bibitem{ultralow_Atin}
\bibinfo{author}{Pal, A.~N.} \& \bibinfo{author}{Ghosh, A.}
\newblock \bibinfo{title}{Ultralow noise field-effect transistor from
  multilayer graphene}.
\newblock \emph{\bibinfo{journal}{Appl. Phys. Lett.}}
  \textbf{\bibinfo{volume}{95}}, \bibinfo{pages}{082105}
  (\bibinfo{year}{2009}).

\bibitem{TransferLength_Pd_Ti}
\bibinfo{author}{Xu, H.} \emph{et~al.}
\newblock \bibinfo{title}{Contact length scaling in graphene field-effect
  transistors}.
\newblock \emph{\bibinfo{journal}{Appl. Phys. Lett.}}
  \textbf{\bibinfo{volume}{100}}, \bibinfo{pages}{103501}
  (\bibinfo{year}{2012}).

\bibitem{VanderPauw}
\bibinfo{author}{Van~der PAUW, L.~J.}
\newblock \bibinfo{title}{A method of measuring the resistivity and hall
  coefficient on lamellae of arbitrary shape}.
\newblock \emph{\bibinfo{journal}{Philips Tech. Rev.}}
  \textbf{\bibinfo{volume}{20}}, \bibinfo{pages}{220--224}
  (\bibinfo{year}{1958}).

\bibitem{bruno2}
\bibinfo{author}{Pellegrini, B.}, \bibinfo{author}{Marconcini, P.},
  \bibinfo{author}{Macucci, M.}, \bibinfo{author}{Fiori, G.} \&
  \bibinfo{author}{Basso, G.}
\newblock \bibinfo{title}{Carrier density dependence of 1/$f$ noise in graphene
  explained as a result of the interplay between band-structure and
  inhomogeneities}.
\newblock \emph{\bibinfo{journal}{J. Stat. Mech. Theor. Exp}}
  \textbf{\bibinfo{volume}{2016}}, \bibinfo{pages}{054017}.

\bibitem{hooge1969contact_noise}
\bibinfo{author}{Hooge, F.} \& \bibinfo{author}{Hoppenbrouwers, A.}
\newblock \bibinfo{title}{Contact noise}.
\newblock \emph{\bibinfo{journal}{Phys. Lett. A}}
  \textbf{\bibinfo{volume}{29}}, \bibinfo{pages}{642--643}
  (\bibinfo{year}{1969}).

\bibitem{macfarlane1950contact_noise_theory}
\bibinfo{author}{MacFarlane, G.}
\newblock \bibinfo{title}{A theory of contact noise in semiconductors}.
\newblock \emph{\bibinfo{journal}{Proc. Phys. Soc. B}}
  \textbf{\bibinfo{volume}{63}}, \bibinfo{pages}{807} (\bibinfo{year}{1950}).

\bibitem{vandamme1974noise_pc}
\bibinfo{author}{Vandamme, L.}
\newblock \bibinfo{title}{1/$f$ noise of point contacts affected by uniform
  films}.
\newblock \emph{\bibinfo{journal}{J. Appl. Phys.}}
  \textbf{\bibinfo{volume}{45}}, \bibinfo{pages}{4563--4565}
  (\bibinfo{year}{1974}).

\bibitem{luo1988theory}
\bibinfo{author}{Luo, M.-Y.}, \bibinfo{author}{Bosman, G.},
  \bibinfo{author}{Van Der~Ziel, A.} \& \bibinfo{author}{Hench, L.~L.}
\newblock \bibinfo{title}{Theory and experiments of 1/$f$ noise in
  schottky-barrier diodes operating in the thermionic-emission mode}.
\newblock \emph{\bibinfo{journal}{IEEE Trans. Electron Devices}}
  \textbf{\bibinfo{volume}{35}}, \bibinfo{pages}{1351--1356}
  (\bibinfo{year}{1988}).

\bibitem{Bozhkov_noise_MS_contacts}
\bibinfo{author}{Bozhkov, V.} \& \bibinfo{author}{Vasiliev, O.}
\newblock \bibinfo{title}{Low-frequency noise in metal-semiconductor contacts
  with local barrier height lowering}.
\newblock \emph{\bibinfo{journal}{Solid-State Electron.}}
  \textbf{\bibinfo{volume}{44}}, \bibinfo{pages}{1487--1494}
  (\bibinfo{year}{2000}).

\bibitem{guttler_noise_Schottky}
\bibinfo{author}{G{\"u}ttler, H.~H.} \& \bibinfo{author}{Werner, J.~H.}
\newblock \bibinfo{title}{Influence of barrier inhomogeneities on noise at
  schottky contacts}.
\newblock \emph{\bibinfo{journal}{Appl. Phys. Lett.}}
  \textbf{\bibinfo{volume}{56}}, \bibinfo{pages}{1113--1115}
  (\bibinfo{year}{1990}).

\bibitem{jayaraman19891}
\bibinfo{author}{Jayaraman, R.} \& \bibinfo{author}{Sodini, C.~G.}
\newblock \bibinfo{title}{A 1/$f$ noise technique to extract the oxide trap
  density near the conduction band edge of silicon}.
\newblock \emph{\bibinfo{journal}{IEEE Trans. Electron Devices}}
  \textbf{\bibinfo{volume}{36}}, \bibinfo{pages}{1773--1782}
  (\bibinfo{year}{1989}).

\bibitem{hung1990unifiedMOSnoise}
\bibinfo{author}{Hung, K.~K.}, \bibinfo{author}{Ko, P.~K.},
  \bibinfo{author}{Hu, C.} \& \bibinfo{author}{Cheng, Y.~C.}
\newblock \bibinfo{title}{A unified model for the flicker noise in
  metal-oxide-semiconductor field-effect transistors}.
\newblock \emph{\bibinfo{journal}{IEEE Trans. Electron Devices}}
  \textbf{\bibinfo{volume}{37}}, \bibinfo{pages}{654--665}
  (\bibinfo{year}{1990}).

\bibitem{Bosman_simulation}
\bibinfo{author}{Hou, F.-C.}, \bibinfo{author}{Bosman, G.} \&
  \bibinfo{author}{Law, M.~E.}
\newblock \bibinfo{title}{Simulation of oxide trapping noise in submicron
  n-channel MOSFETs}.
\newblock \emph{\bibinfo{journal}{IEEE Trans. Electron Devices}}
  \textbf{\bibinfo{volume}{50}}, \bibinfo{pages}{846--852}
  (\bibinfo{year}{2003}).

\bibitem{JDeen_MOS2}
\bibinfo{author}{Xie, X.} \emph{et~al.}
\newblock \bibinfo{title}{Low-frequency noise in bilayer {MoS}$_2$ transistor}.
\newblock \emph{\bibinfo{journal}{ACS Nano}} \textbf{\bibinfo{volume}{8}},
  \bibinfo{pages}{5633--5640} (\bibinfo{year}{2014}).

\bibitem{low_noise_SiP}
\bibinfo{author}{Shamim, S.}, \bibinfo{author}{Weber, B.},
  \bibinfo{author}{Thompson, D.~W.}, \bibinfo{author}{Simmons, M.~Y.} \&
  \bibinfo{author}{Ghosh, A.}
\newblock \bibinfo{title}{Ultralow-noise atomic-scale structures for quantum
  circuitry in silicon}.
\newblock \emph{\bibinfo{journal}{Nano Lett.}} \textbf{\bibinfo{volume}{16}},
  \bibinfo{pages}{5779--5784} (\bibinfo{year}{2016}).

\bibitem{Tersoff_Avouris_CNT}
\bibinfo{author}{Heinze, S.} \emph{et~al.}
\newblock \bibinfo{title}{Carbon nanotubes as schottky barrier transistors}.
\newblock \emph{\bibinfo{journal}{Phys. Rev. Lett.}}
  \textbf{\bibinfo{volume}{89}}, \bibinfo{pages}{106801}
  (\bibinfo{year}{2002}).

\bibitem{vidya_GBnoise}
\bibinfo{author}{Kochat, V.} \emph{et~al.}
\newblock \bibinfo{title}{Magnitude and origin of electrical noise at
  individual grain boundaries in graphene}.
\newblock \emph{\bibinfo{journal}{Nano Lett.}} \textbf{\bibinfo{volume}{16}},
  \bibinfo{pages}{562--567} (\bibinfo{year}{2015}).

\bibitem{deanBN}
\bibinfo{author}{Dean, C.} \emph{et~al.}
\newblock \bibinfo{title}{Boron nitride substrates for high-quality graphene
  electronics}.
\newblock \emph{\bibinfo{journal}{Nat. Nanotechnol.}}
  \textbf{\bibinfo{volume}{5}}, \bibinfo{pages}{722--726}
  (\bibinfo{year}{2010}).

\bibitem{ZomerAPL2011}
\bibinfo{author}{Zomer, P.~J.}, \bibinfo{author}{Dash, S.~P.},
  \bibinfo{author}{Tombros, N.} \& \bibinfo{author}{van Wees, B.~J.}
\newblock \bibinfo{title}{A transfer technique for high mobility graphene
  devices on commercially available hexagonal boron nitride}.
\newblock \emph{\bibinfo{journal}{Appl. Phys. Lett.}}
  \textbf{\bibinfo{volume}{99}}, \bibinfo{pages}{232104}
  (\bibinfo{year}{2011}).

\bibitem{ArindamArxiv}
\bibinfo{author}{{Ghosh}, A.}, \bibinfo{author}{{Kar}, S.},
  \bibinfo{author}{{Bid}, A.} \& \bibinfo{author}{{Raychaudhuri}, A.~K.}
\newblock \bibinfo{title}{{A set-up for measurement of low frequency
  conductance fluctuation (noise) using digital signal processing techniques}}.
\newblock \emph{\bibinfo{journal}{Preprint at
  https://arxiv.org/abs/cond-mat/0402130}}  (\bibinfo{year}{2004}).

\end{thebibliography}
\end{document}